# Empirical Evaluations of Preprocessing Parameters' Impact on Predictive Coding's Effectiveness

Rishi Chhatwal, Esq.
Legal
AT&T Services, Inc.
Washington, D.C. USA
Email: rc134a@att.com

Nathaniel Huber-Fliflet
Legal Technology Solutions
Navigant Consulting, Inc.
Washington, D.C. USA
Email: nathaniel.huber-fliflet@navigant.com

Robert Keeling, Esq.
Complex Commercial Litigation
Sidley Austin LLP
Washington, D.C. USA
Email: rkeeling@sidley.com

Dr. Jianping Zhang
Legal Technology Solutions
Navigant Consulting, Inc.
Washington, D.C. USA
Email: jianping.zhang@navigant.com

Dr. Haozhen Zhao
Legal Technology Solutions
Navigant Consulting, Inc.
Washington, D.C. USA
Email: haozhen.zhao@navigant.com

*ABSTRACT* – Predictive coding, once used in only a small fraction of legal and business matters, is now widely deployed to quickly cull through increasingly vast amounts of data and reduce the need for costly and inefficient human document review. Previously, the sole front-end input used to create a predictive model was the exemplar documents (training data) chosen by subject-matter experts. Many predictive coding tools require users to rely on static preprocessing parameters and a single machine learning algorithm to develop the predictive model. Little research has been published discussing the impact preprocessing parameters and learning algorithms have on the effectiveness of the technology. A deeper dive into the generation of a predictive model shows that the settings and algorithm can have a strong effect on the accuracy and efficacy of a predictive coding tool. Understanding how these input parameters affect the output will empower legal teams with the information they need to implement predictive coding as efficiently and effectively as possible. This paper outlines different preprocessing parameters and algorithms as applied to multiple real-world data sets to understand the influence of various approaches.

*Keywords – predictive coding, technology assisted review, electronic discovery, ediscovery, e-discovery*

## I. INTRODUCTION

Information management has become a significant business challenge, with the global volume of electronically stored information growing at a rapid pace (doubling roughly three times since 2010 [1]). Companies regularly spend millions of dollars producing responsive electronically stored documents for litigation matters [2]. The review process generates the bulk of e-discovery costs [2]. Predictive coding is frequently applied to legal matters with millions of documents that require attorney review. For example, one of our current matters contains more than 21 million documents that need to be classified as relevant or not relevant. To more efficiently cull through massive volumes of data for relevant information, companies turn to predictive coding, also referred to as technology assisted review, or text categorization. Predictive coding applies a supervised machine learning algorithm to build a predictive model to automatically classify documents into predefined categories of interest, such as relevant and not relevant or privileged and not privileged.

Predictive coding, already highly valued in litigation settings, is increasingly being embraced in other legal matters such as Department of Justice requests, and mergers and acquisitions. As its use increases, the technical side of predictive coding – effectively selecting the preprocessing parameters and learning algorithms – largely remains an enigma to those involved in e-discovery.

The long-standing presumption is that the accuracy and effectiveness of the predictive coding process relies heavily on the exemplar documents used to build the model. Our experiments show otherwise. This paper demonstrates that adjusting the preprocessing parameters required to implement the technology – an area that is traditionally overlooked – also has a dramatic impact on results.

Choosing ineffective parameter combinations to build a predictive model can result in missing important, case sensitive documents and lost cost savings – in some cases, precision was reduced by more than 34%, reducing cost savings by more than 60%. Our experiments, which comprised three data sets from real legal matters across several industries, generated predictive models using various combinations of preprocessing parameter settings to determine the best overall combination for predictive coding effectiveness. In this paper, we (i) outline predictive coding and introduce different types of preprocessing parameters and machine learning algorithms; (ii) describe the experiments and the data sets used; and (iii) report our results and findings, highlighting key components that have the largest influence on results.

## II. PREDICTIVE CODING

The challenge posed by this ocean of information is especially acute when the company is asked to wade through it to respond to e-discovery requests in litigation. Companies regularly spend millions of dollars in the production of relevant electronically stored documents [2]. Most of these costs are not associated with the collection or processing of the data, but are rather incurred when the documents are reviewed [2].

The traditional approach to document review is quickly becoming less of an option in today's legal environment. Previously, keyword searches and human reviewers were enough to constitute a comprehensive and defensible review strategy; however, increasing document volumes, complexity and costs require the traditional approach to evolve. Predictive coding has long been recognized for offering significant benefits to companies needing to cull through massive volumes of data to find relevant information. Predictive coding applies advanced machine learning techniques to the text of documents to automatically classify un-reviewed documents into predefined categories of interest, such as relevance or privilege. The classification models are trained through supervised learning – meaning the model is built from a human-reviewed subset of documents. Legal teams can then leverage the results to review the documents that are most likely relevant so they can more quickly understand the content within the document population. It also allows legal teams to consider excluding likely irrelevant documents from review or shifting the review responsibilities of those likely irrelevant documents to lower cost review teams.

By leveraging predictive coding, a company may significantly reduce the time and cost associated with the e-discovery process. Predictive coding does not eliminate human review, rather, it sorts through massive volumes of data to reduce the amount of documents that require human review. Generally, like traditional reviews, stores of electronic information must still be collected and processed before predictive coding is applied. Unlike traditional reviews, however, where humans do the bulk of the work, under predictive coding humans provide exemplars of relevant documents and validate the machine predictions in a kind of "code, rinse and repeat" cycle until an acceptable level of confidence in the result is obtained.

Because predictive coding has been so effective at identifying more documents like the exemplars, most of the predictive coding debate has centered on identifying and verifying the appropriate training exemplars. However, not all predictive coding is equal. Realizing the full benefits of predictive coding requires a keen understanding of how to establish the preprocessing parameters and train the algorithms. As this paper highlights, small changes in the process can reap considerable savings.

## III. PREPROCESSING PARAMETERS

Preprocessing, an important component of predictive coding and required to develop a predictive model, transforms a document into a vector of feature values and selects a supervised machine learning algorithm for predictive model development. Our tool provided the user with the ability to select a machine learning algorithm and to tune many preprocessing parameters to achieve optimized results for a given predictive coding project. The choice of preprocessing parameters and machine learning algorithm can have a significant impact on the results of the predictive model. Our study analyzed various implementations of the following preprocessing parameters and machine learning algorithms:

- Word Stemming
- N-Grams
- Token Value Type
- Number of Tokens
- Down Sampling
- Support Vector Machine
- Logistic Regression

To begin, we use the bag-of-words approach to represent a document as a vector of feature values (numerical representation of a document). The text (e.g., sentence, document) is represented as a 'bag' of all its words, disregarding grammar and word order but keeping track of repeated words. Each unique word (or n-gram), referred to as a 'token', is considered as a feature. This approach allows for a simplified representation of the text within the document population.

The *Word Stemming* parameter defines whether stemming is applied to any documents in the corpus. We used Porter's stemming algorithm for this study.

An *N-Gram* is a contiguous sequence of *n* tokens from the text of a document. The n-grams parameter takes a positive integer as its value. When the n-grams parameter is *n*, all 1-grams (one word), 2-grams (two words), 3-grams (three words), and so on, are generated as tokens to represent a document. The n-grams parameter provides an opportunity to evaluate the impact the combination of words has in defining the category (e.g., relevance, privilege). For example, independently, words like 'white' and 'house' have a very different meaning than 'White House'. The same is true for 'breast' and 'cancer' vs. 'Breast Cancer' and the n-grams parameter provides an opportunity to take this into account. N-grams are established for all documents in the corpus.

A token or an n-gram is considered as a feature, which takes a value. For experimentation, we chose to implement four different types of *Token Values*: binary, term frequency, normalized term frequency, and term frequency-inverse document frequency.

A binary token value is the most popular – the token either exists in the document or it does not. If a token occurs in a document, it is 1; otherwise, it is 0.

The second type is term frequency, which takes an integer larger than or equal to 0. Its value is the number of times the token occurs in a document. Sometimes frequency can indicate its relevance. For example, an article that mentions 'George Washington' once may or may not be about George Washington. However, an article that mentions 'George Washington' 15 times is more likely to focus on the president.

The third type is normalized (augmented) term frequency. It is computed using the following formula:

$$NTF(t,d) = 0.5 + 0.5 \times \frac{TR(t,d)}{Max\{TR(t_i,d)\}} \quad (1)$$

NTF(t,d) is the normalized term frequency for term *t* in document *d*.

- TR(t, d) is the term frequency for *t* in *d*,
- and Max{TR($t_i$, d)} is the maximum term frequency for all terms in d.

Not all frequently used words or tokens are effective at defining the category, so normalized term frequency helps ensure that less frequently occurring words are not overshadowed by frequently occurring words. For example, a Navigant press release may mention 'Navigant' many times, but that does not mean the press release is only about 'Navigant'. The press release may focus on quarterly earnings but use the phrase 'quarterly earnings' fewer times than 'Navigant'. In this example, 'quarterly earnings' is more effective at distinguishing the press release's content than 'Navigant'.

To further illustrate normalized term frequency's function, assume that 'Navigant' is the most frequent token in a document, occurring 10 times and within the same document, 'quarterly earnings' occurs twice. Using the term frequency value type, 'Navigant' is considered five times more important than 'quarterly earnings' ((10/2) = 5). However, using normalized term frequency, the value of 'Navigant' is: 0.5 + 0.5*(10/10) = 1 and the value of 'quarterly earnings' is: 0.5 + 0.5*(2/10) = 0.6. 'Navigant' is still more important, but not five times more important.

The last token value type is term frequency-inverse document frequency (TFIDF*)*, which is a value that intends to reflect how important a given token is to an individual document in the document population. The TFDIF value compares a token's frequency to its uniqueness in the document population. If a token is a very common, non-stop word like, 'time' or 'day' and occurs across many documents in the population, it should have less impact on the model than a token that occurs frequently in one document.

The following formula is used to determine TFIDF:

$$TFIDF(t,d) = NTF(t,d) \times \log\frac{N}{N_t} \quad (2)$$

- NTF(t,d) is defined in the normalized term frequency section above,
- and *N* is the number of documents in the collection,
- and $N_t$ is the number of documents including the term in the collection of documents.

Training documents may contain millions of different words, many of which are irrelevant to the predictive coding exercise, adding noise to the process and reducing the effectiveness of the machine learning algorithm. We use Information Gain, a feature (token) selection algorithm to select a subset of the most effective tokens to build a predictive model. The information gain of a given token is generally based on the token's effectiveness at discriminating between the categories of interest – the higher the discrimination power, the higher the information gain. Training the model with tokens that are most effective at defining the relevant and not relevant classes will reduce the noise that irrelevant tokens create. A number of studies have confirmed the effectiveness of information gain as a token selection criterion for predictive modeling tasks [3].

With the information gain of each token in the training set established, the most effective number of tokens can be targeted and selected. The *Number of Tokens* parameter simply defines the number of top most discriminating tokens to use from the training set. Combining the results of information gain and an optimized number of tokens together transforms the available tokens in the training set into a narrow and highly discriminant set of tokens for modeling.

The distribution of the modeling category (e.g., between relevance and non-relevance, privilege and non-privilege) is often unbalanced within the document corpus of a legal matter. In an unbalanced data set, the majority class (usually not relevant documents) is represented by a large percentage of all the documents, while the other, minority class (usually relevant documents), has only a small percentage of all documents. Studies [4] have shown that unbalanced class distributions result in poor performance using many machine learning algorithms. Down sampling is a frequently used approach to address the challenges caused by unbalanced class distribution. Instead of using the entire set of negative (majority class) training examples, a subset of negative examples is selected, such that the resulting training data is less unbalanced and recall may be enhanced. The *Down Sampling* parameter defines the percentage of negative (e.g., not relevant) training documents used to build a model.

There are many different machine learning algorithms that can be used for text categorization including, Support Vector Machines, Logistic Regression, Naïve Bayesian Classifier, and Convolutional neural networks. For our study, we chose to evaluate two popular machine learning algorithms, Support Vector Machines (SVM) and Logistic Regression (LR). SVM has been widely used to develop predictive models [5].

## IV. EXPERIMENTAL DATA SETS AND DESIGN

In this section, we describe the data sets used in our experiments and the experiment setup. Our experimentation was designed to thoroughly evaluate the impact of important preprocessing parameters on the effectiveness of predictive models.

### A. DATA SETS

Real legal matters from three different industries comprise our three data sets. The predictive coding task was to identify all relevant documents. Each data set contained email, Microsoft Office documents, and other text-type documents, consisting of a set of training documents and a set of validation documents used to calculate the models' recalls and precisions. Attorneys manually reviewed documents in both data sets to confirm relevance decisions. The documents within each validation set were randomly selected from the entire corpus of the specific data set. Table 1 details the document statistics of each data set. Project 1 and Project 2 have an unbalanced class distribution, although their training sets are not as unbalanced. Documents in Project 3 are evenly distributed among relevant and not relevant.

Table 1. Data set statistics

| Document Class Distribution | Project 1 | Project 2 | Project 3 |
|---|---|---|---|
| Training - Relevant | 1,126 | 527 | 5,743 |
| Training - Not Relevant | 2,897 | 1,114 | 6,540 |
| Validation - Relevant | 206 | 292 | 801 |
| Validation - Not Relevant | 1,368 | 1,298 | 788 |

### B. EXPERIMENTAL SETUP

16,800 experiments were performed for this study using various combinations of preprocessing parameter values and machine learning algorithms. Table 2 details the experimental values for each parameter.

Table 2. Parameters and values

| Parameters | Parameter Values |
|---|---|
| Word Stemming | Yes, No |
| N-Grams | 1, 2, 3, 4 |
| Token Value Type | Binary, Frequency, Normalized Term Frequency, TFIDF |
| Number of Tokens | 1,000, 3,000, 5,000, 7,000 10,000, 15,000, 20,000, 25,000, 30,000, 35,000, 40,000, 45,000, 50,000 |
| Down Sampling | 25%, 50%, 75%, 100% |
| Machine Learning Algorithm | Support Vector Machine (SVM), Logistic Regression (LR) |

Both the SVM and LR machine learning algorithms and their default parameter settings were selected from LibLinear, an open source library for large-scale linear classification. The linear kernel was used for SVM.

In each experiment, the training documents from each project generated 5,600 predictive models to test all combinations of the parameters listed in Table 2; results were evaluated with the projects' corresponding validation set. The performance of each experimental model was analyzed using recall, precision, and the percentage of documents requiring attorney review as performance metrics.

The results of each parameter setting's impact was calculated using the average of all other parameter settings' precisions and percentages of documents requiring review at a specific recall rate. Using precision as an example, Project 3 generated 2,800 models using SVM and all other combinations of parameter settings (the SVM Model Experiments) and generated an additional 2,800 using LR and the same combination of parameter settings used for SVM (the LR Model Experiments). To compare the overall performance of SVM versus LR, first an average percentage of documents reviewed was used to characterize each model's performance, which was calculated by averaging the percentage of the documents reviewed at specific recall rates (30%, 40%, 50%, 60%, 70%, 80%, and 90%) for that model. Then for the two sets of experiments, we determined the average percentages of documents reviewed for each set of experiments.

## V. EXPERIMENTAL RESULTS

In this section, we report and discuss our experimental results for Token Value Types, Learning Algorithms, and Down Sampling. For each parameter, we report the average percentages of documents requiring review in order to achieve the corresponding recalls as described in section 4.2 Experimental Setup. In other words, we report which parameter is most successful at reducing the number of not relevant documents requiring review. The averages were calculated using the results of 5,600 predictive models generated for each project using the different combinations of parameter settings. Comprehensive results for Word Stemming, N-Grams, and Number of Tokens are excluded from this paper to keep within the paper's length requirement.

### A. TOKEN VALUE TYPES

Figure 1 displays the average percentages of documents reviewed for the four different token value types: binary, frequency, normalized term frequency, and TFIDF.

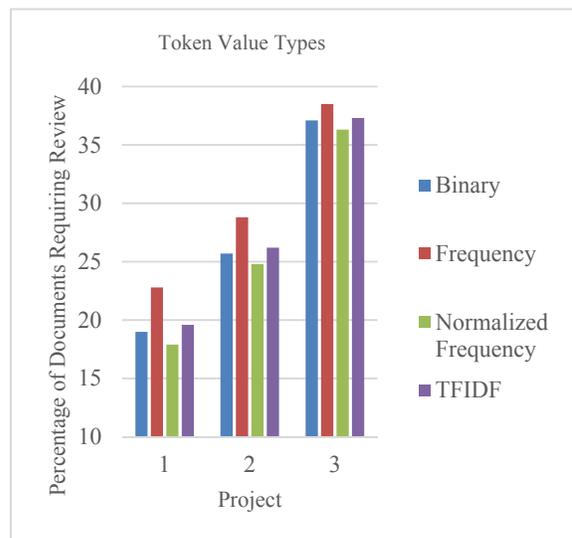

Figure 1. Token Value Types

For all three projects, normalized term frequency performed the best, minimizing the percentage of documents requiring review. Frequency required review of the most documents, with binary requiring less documents than TFIDF.

Both binary and TFIDF are widely used to create predictive models [6]. However, TFIDF was developed for information retrieval and is not necessarily effective for predictive modeling. A term appearing in very few documents does not mean it is effective at distinguishing between one class of documents and another. For example, a term occurring in two documents, one relevant and one not relevant, is not effective at identifying relevant documents.

Predictive models generated using binary, the second best performing value, would require review of 1.1%, 0.9%, and 0.8% more documents for Projects 1, 2, and 3, respectively, when compared to normalized term frequency. Now consider a legal matter with one million documents requiring review: even one percent inefficiency would result in 10,000 extra documents for review.

### B. MACHINE LEARNING ALGORITHMS

We conducted experiments to compare two popular machine learning algorithms, Support Vector Machine (SVM) and Logistic Regression (LR). Figure 2 displays the average percentages of documents requiring review for the results of the 5,600 experimental models generated for the three projects using the two machine learning algorithms.

Figure 2. Machine Learning Algorithms

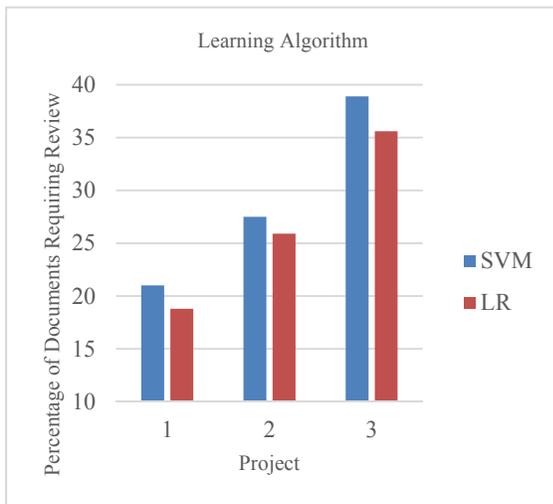

SVM is considered to be one of the best performing learning algorithms for text categorization [7], but here our results show that LR achieved better performance on all three projects. This is true across all recall rates.

Predictive models generated using SVM would require review of 2.2%, 1.6%, and 3.3% more documents for Projects 1, 2, and 3, respectively, when compared to LR. Using the one million document legal matter example, this would require reviewing an extra 16,000 to 33,000 documents.

### C. DOWN SAMPLING

Average precisions for different down sampling values did not vary greatly across the three projects, but the differences in the average percentages of the documents requiring review for Project 1 were significant. Figures 3, 4, and 5 display the average percentages of documents requiring review for the three projects at different down sampling percentages for different relevant recall rates. The results show that down sampling significantly improves performance at higher recall levels, but reduces performance at lower recalls for Project 1 and 2. This is expected for two reasons: the class distribution for these two projects is unbalanced and fewer negative examples allow the learning algorithm to generate a model that is biased toward the positive class. For Project 3, down sampling negatively affects performance (other than at 90% recall) because the class distribution for it is roughly even.

The real world impact of down sampling on Project 1 is compelling. A model generated with all the not relevant training documents would require review of 4.6% more documents than a model using 25% down sampling (25% of the original not relevant training documents). Considering the one million document legal matter example a final time, this would result in 46,000 extra documents for review.

Figure 3. Down Sampling for Project 1

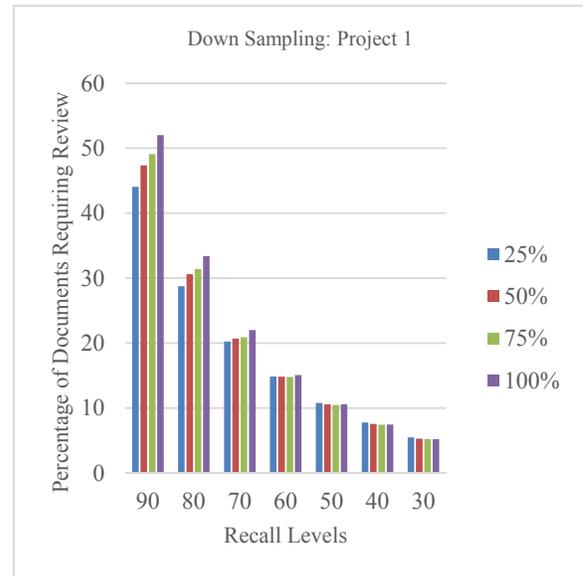

Figure 4. Down Sampling for Project 2

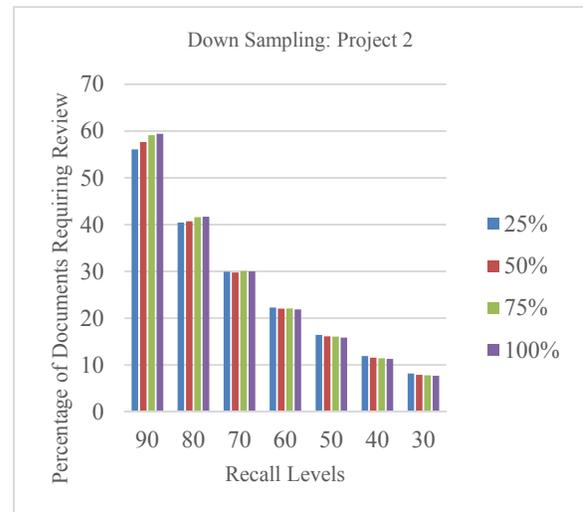

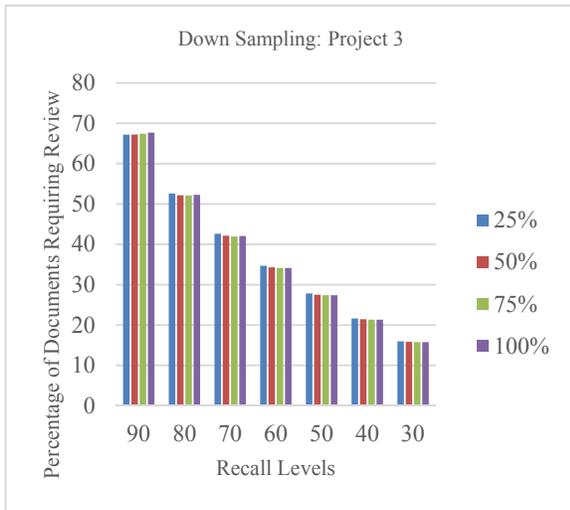

Figure 5. Down Sampling for Project 3

## VI. CONCLUSIONS

Predictive coding is a black box to most legal teams. At best, they may know the machine learning algorithm, but the preprocessing parameter settings used to generate a predictive model are largely a mystery.

This study demonstrates that small changes in the preprocessing parameters and the machine learning algorithm choice can have a significant impact on the results of a predictive model.

We assumed that the preprocessing parameters had some level of impact on the quality of a predictive model's results. We tested this theory using an empirical analysis of five preprocessing parameters and two machine learning algorithms, performing nearly 17,000 experiments. The results showed:

- The normalized term frequency token value achieved better results than binary, TFIDF, and frequency. Binary, the second best performing token value, would require review of 0.9% more documents when compared to normalized term frequency.

- The Logistic Regression algorithm performed much better than the Support Vector Machine algorithm. On Project 3, LR could exclude 3.3% more documents from review.

- Down sampling performed well when attempting to achieve high recalls on data sets with unbalanced class distributions. Project 1 generated a model requiring review of 4.64% less documents at 80% recall when 25% down sampling was applied, compared to a model that used all available not relevant training documents. However, down sampling did not perform well at low recalls and on an evenly distributed data set. The results suggest that down sampling should never be used to target relevant documents; only to drive up the recall on an unbalance data set.

Weak combinations of preprocessing parameters and the machine learning algorithm choice have a dramatic impact on the results of the model. Figure 6 displays the results of the best and worst performing model for Project 1 developed using all preprocessing parameter settings. The strongest combination of results would be 34.62% more precise and would reduce the volume of review by 60.61% when compared to the worst combination of parameters and algorithm.

Figure 6. Strongest and Weakest Combination of Preprocessing Parameters

| Parameter Type | Strongest | Weakest |
|---|---|---|
| Word Stemming | No | Yes |
| Number of Tokens | 25,000 | 7,000 |
| N-Grams | 1 | 4 |
| Down Sampling | 25% | 100% |
| Token Value Type | Normalized Term Frequency | TFIDF |
| Machine Learning Algorithm | LR | SVM |
| **Precision @ 80% Recall:** | 47.28 | 12.66 |
| **Documents Requiring Review @ 80% Recall:** | 22.17 | 82.78 |

Our experiments suggest that, on average, the best performing combination of preprocessing parameters and machine learning algorithm to generate a predictive model for a legal matter are:

- Logistic Regression,
- at least 10,000 tokens,
- 1-Gram,
- normalized term frequency,
- with word stemming turned off,
- and down sampling used when needed.

The training process described throughout this paper used simple passive learning – a training set is established and the model is considered final once it is generated. When using active or ensemble learning, the document classification task must repeat many times to develop the best performing model and requires classifying all documents in the data set repeatedly. With continuous active learning, the predictive model's training set is updated regularly, often daily, as more training documents become available. For example, we have a project leveraging continuous active learning that now has more than 392,000 training documents. The process to classify documents and develop predictive models using large training sets is very time consuming and big data technologies could dramatically speed up these tasks. In the future, we plan to test additional big data methodologies and technology to improve the speed of cumbersome modeling tasks.